\documentclass[
 reprint,
 amsmath,amssymb,
 nofootinbib,
 aps,
 prd,
 superscriptaddress,
 floatfix,
]{revtex4-1}
\usepackage{aas_macros}
\usepackage{graphicx}
\usepackage{dcolumn}
\usepackage{bm}
\usepackage{amsmath}
\usepackage{float}
\usepackage{lipsum}
\usepackage{xcolor}
\usepackage{textcomp}
\usepackage{latexsym}
\usepackage[urlcolor=blue, colorlinks=true, linkcolor=black, citecolor = black]{hyperref}
\usepackage{cleveref}[2013/12/23]
\usepackage{mathtools}
\usepackage{url}
\usepackage{comment}
\usepackage[utf8]{inputenc}
\usepackage[normalem]{ulem}
\usepackage{xspace}
\usepackage{acronym}

\newcommand{\beq}{\begin{equation}}
\newcommand{\eeq}{\end{equation}}

\newcommand{\bw}{\texttt{BayesWave}\xspace}
\newcommand{\bl}{\texttt{BayesLine}\xspace}
\newcommand{\bilby}{\texttt{Bilby}\xspace}
\newcommand{\LIGOlabMIT}{\affiliation{LIGO Laboratory, Massachusetts Institute of Technology, 185 Albany St, Cambridge, MA 02139, USA}}
\newcommand{\MKI}{\affiliation{Department of Physics and Kavli Institute for Astrophysics and Space Research, Massachusetts Institute of Technology, 77 Massachusetts Ave, Cambridge, MA 02139, USA}}
\newcommand{\Imperial}{\affiliation{Department of Physics, Imperial College London, London, UK, SW7 2AZ}}

\begin{document}

\title{Quantifying the Effect of Power Spectral Density Uncertainty on Gravitational-Wave Parameter Estimation for Compact Binary Sources}

\author{Sylvia Biscoveanu} \LIGOlabMIT \MKI
\author{Carl-Johan Haster} \LIGOlabMIT \MKI
\author{Salvatore Vitale} \LIGOlabMIT \MKI
\author{Jonathan Davies} \Imperial

\begin{abstract}
In order to perform Bayesian parameter estimation to infer the source properties of gravitational waves from \acp{CBC}, the noise characteristics of the detector must be understood. 
It is typically assumed that the detector noise is stationary and Gaussian, characterized by a \acf{PSD} that is measured with infinite precision. 
We present a new method to incorporate the uncertainty in the power spectral density estimation into the Bayesian inference of the binary source parameters
and apply it to the first 11 \ac{CBC} detections reported by the LIGO-Virgo Collaboration. We find that incorporating the \ac{PSD} uncertainty only leads to variations in the positions and widths of the binary parameter posteriors on the order of a few percent. Our results are publicly available for download on git~\cite{data_release}.

\end{abstract}

\maketitle

\acrodef{PSD}[PSD]{power spectral density}
\acrodef{ASD}[ASD]{amplitude spectral density}
\acrodef{SNR}[SNR]{signal-to-noise ratio}
\acrodef{BH}[BH]{black hole}
\acrodef{BBH}[BBH]{binary black hole}
\acrodef{BNS}[BNS]{binary neutron star}
\acrodef{NS}[NS]{neutron star}
\acrodef{BHNS}[BHNS]{black hole--neutron star binaries}
\acrodef{NSBH}[NSBH]{neutron star--black hole binary}
\acrodef{CBC}[CBC]{compact binary coalescence}
\acrodef{GW}[GW]{gravitational wave}
\acrodef{PDF}[PDF]{probability density function}
\acrodef{PE}[PE]{parameter estimation}
\acrodef{CL}[CL]{credible level}
\acrodef{IFO}[IFO]{interferometer}
\acrodef{MCMC}[MCMC]{Markov Chain Monte Carlo}

\section{Introduction} \label{sec:intro}
A gravitational wave detector, such as the ground-based laser interferometers LIGO and Virgo~\cite{TheLIGOScientific:2014jea,TheVirgo:2014hva}, is assumed to generate an output consisting of background Gaussian noise. 
Gravitational-wave signals, as well as non-Gaussian noise transients, will introduce a deviation of the detector output from the baseline noise behavior. 
As the sensitivity of the network of ground-based laser interferometers searching for gravitational waves improves~\cite{Aasi:2013wya}, so too must our understanding of their noise properties.
These noise properties are usually characterized by the \acf{PSD} in each detector, which is a required input for both low latency searches for gravitational waves and further source characterization via Bayesian parameter estimation~\cite{Messick:2016aqy, Sachdev:2019vvd, Nitz:2018rgo, Usman:2015kfa, Adams:2015ulm, Chu:2017, Veitch:2014wba, Ashton:2018jfp, Huang:2020ysn}. 
Both of these types of analyses typically require that the \ac{PSD} is measured with infinite precision, and so far LIGO/Virgo template-based results have not accounted for the uncertainty in the \ac{PSD} estimation since a single point estimate has been used for each analysis segment~\cite{LIGOScientific:2018mvr}. Unmodeled searches and follow-up analyses, including \ac{CBC} waveform reconstruction~\cite{LIGOScientific:2018mvr} and short-duration gravitational-wave transient searches~\cite{Abbott:2019prv} have, however, included marginalization over the uncertainty in the \ac{PSD} estimation. 
While the  low-latency searches employed by LIGO account for the variability of the PSD on longer timescales by recalculating it periodically~\cite{Nitz:2018rgo, Sachdev:2019vvd}, all \ac{PSD} estimation methods formally assume that the detector noise is Gaussian and stationary, such that the noise properties do not change over the course of the data segment used in the calculation~\cite{Chatziioannou:2019zvs}. 
However, these assumptions are not generally true and can impact the sensitivity of the searches when the noise is mis-characterized~\cite{TheLIGOScientific:2017lwt}. 

The properties of the noise do vary in time, though usually the stationarity timescale is much longer than the analysis segment for transient gravitational-wave signals~\cite{Aasi:2013jjl,Chatziioannou:2019zvs}. 
Recently, a new method for relaxing the stationarity assumption was proposed and applied in \cite{Zackay:2019kkv} and \cite{Venumadhav:2019tad}, respectively, where the non-stationarity of the data over the duration of the segment used to calculate the \ac{PSD} is accounted for by applying a ``drift'' correction to the \ac{PSD} obtained by tracking the time-dependent variance of the overlaps between the data and signal templates used in low-latency searches. They demonstrated that this had a significant impact on the recovered trigger distribution for the search pipeline. A similar method to account for the change of the detector noise properties over time via the dynamic renormalization of the search trigger ranking statistic was developed in \cite{Mozzon:2020gwa} and applied in \cite{Nitz:2019hdf}, leading to an improvement in the search sensitivity for low-mass compact binary systems.
In addition to these slow variations in the \ac{PSD}, the data is often plagued by short (ms) transient non-Gaussian excursions, known as glitches~\cite{Nuttall:2015dqa, Zevin:2016qwy,TheLIGOScientific:2016zmo}. 
In low latency, glitches are typically excised from the data using a procedure known as ``gating''~\cite{TheLIGOScientific:2017lwt}, while in higher latency they can be modeled and subtracted from the data, as was the case for the glitch present in the Livingston detector during the first \ac{BNS} merger, GW170817~\cite{Pankow:2018qpo}. 

Finally, the noise power spectral density cannot be measured with infinite precision and must be estimated from the data itself in one of two ways.
The first is a modification of the standard Welch's method~\cite{Welch:1967}, in which a longer stretch of data either before or after, but always excluding, the analysis segment is divided into smaller segments with the same duration as the analysis segment, and the resulting \ac{PSD} is the mean of the periodogram for each of these sub-segments. 
This is known as an ``off-source'' method since the data used to compute the \ac{PSD} excludes the analysis segment.
This requires the noise to be stationary over the entire segment used for the \ac{PSD} estimation, which can be on the order of $~1000~\mathrm{s}$. 
Because glitches can bias the mean, the median periodogram is generally used in gravitational-wave data analysis instead~\cite{Allen:2005fk, Veitch:2014wba}. 

The second, or ``on-source'' method only uses the data from the specific segment under analysis, therefore assumed to contain both Gaussian noise and a non-Gaussian signal component.
The Gaussian contribution is inferred from the data and represented as a frequency-dependent noise variance parameterized in terms of a phenomenological model with two separate components, a cubic spline describing the broadband Gaussian process assumed to generate the noise itself combined with a set of Lorenztians describing narrow-band features\footnote{The narrowband features can typically be attributed either to the resonances of the cables suspending the test masses, known as ``violin modes'', the AC electrical supply ``power line'', or the ``calibration lines'' which are injected into the data by driving the test masses at known frequencies~\cite{Littenberg:2014oda, Driggers:2018gii, Davis:2018yrz, Vajente:2019ycy}.}.
The number and position of both the spline points and the Lorenztians, as well as the line widths and amplitudes, are themselves free parameters in the models that are explored through a trans-dimensional \ac{MCMC} algorithm~\cite{Littenberg:2014oda}.
The non-Gaussianity due to the presence of the astrophysical signal is modeled using sine-Gaussian wavelets~\cite{Littenberg:2014oda, Cornish:2014kda}. 
Both the on-source and off-source methods assume the noise to be stationary and Gaussian, but as the off-source method requires stationarity over a duration more than an order of magnitude longer than the on-source method, this assumption is more likely to hold true for the on-source method~\cite{Chatziioannou:2019zvs}.

In this paper, we demonstrate a new method to relax the assumption that the \ac{PSD} measurement is infinitely precise as applied to Bayesian parameter estimation for gravitational waves from compact binaries, where instead of using a single point estimate for the \ac{PSD}, we marginalize over the uncertainty in the \ac{PSD} estimation. 
Other studies have previously looked at the effects of the methods and uncertainty associated with modeling the noise \acl{PSD} on compact binary parameter estimation. 
\cite{Aasi:2013jjl} found that using different noise realizations and hence different \acp{PSD} has a similar impact on the variation in the recovered source parameters as using different waveform models. 
The idea of marginalizing over the uncertainty in the \ac{PSD} was first proposed in \cite{Rover:2008yp} and \cite{Rover:2011qd} by analytically marginalizing the standard Gaussian likelihood for gravitational wave data over the uncertainty in the \ac{PSD} and arriving at the Student's T likelihood. This technique was employed in \cite{Banagiri:2019lon} to obtain unbiased parameter estimates for the properties of neutron star postmerger remnants in the context of the millisecond magnetar model.
\cite{Smith:2017vfk} proposed a similar method using a Gaussian prior on the \ac{PSD} instead of a scaled inverse $\chi^{2}$-distribution for compact binary parameter estimation in the presence of a Gaussian stochastic background.  
In \cite{Veitch:2014wba} and \cite{Littenberg:2013gja} the uncertainty in the \ac{PSD} was parameterized as a scale factor that modifies the point estimate for a fixed number of frequency segments, which led to significant improvements of the consistency of compact binary parameter estimation results in real LIGO data.
Methods for simultaneously measuring the \ac{PSD} using a different parameterization in the presence of an astrophysical signal were developed in \cite{Littenberg:2014oda} and \cite{Cornish:2014kda}, although their signal model is a sum of wavelets and not a 17-dimensional compact binary waveform. Another method to simultaneously estimate the \ac{PSD} and astrophysical signal parameters was presented in \cite{Edwards:2015eka}, which used a nonparametric approach to model the \ac{PSD} in the presence of a gravitational-wave burst from core-collapse supernovae. 
More recently, \cite{Chatziioannou:2019zvs} investigated the differences between the two methods for computing the \ac{PSD} outlined above and found that the on-source method provides a better agreement with the statistical assumptions about the data described previously. 

With this in mind, our method differs from the one proposed in~\cite{Rover:2008yp} because the analytic marginalization still requires the \ac{PSD} point estimate to be computed via the off-source method, while our method uses the full \ac{PSD} posterior calculated via the on-source method, recovering many of the same advantages that are detailed in~\cite{Chatziioannou:2019zvs}. 
Additionally, while the parameterization in \cite{Veitch:2014wba} and \cite{Littenberg:2013gja} models the scale factor as constant over some range of frequencies, the posteriors we obtain for the \ac{PSD} using the on-source method allow for variation at much higher frequency resolution.

We note that a similar method was proposed in~\cite{Ashton:2019leq} in the context of marginalizing over the uncertainty due to the choice of waveform model, although the implementation differs from the method presented in this work since the waveform model is not an independent parameter for which posteriors are obtained, unlike the \ac{PSD}.

The rest of this paper is organized as follows. 
In Section~\ref{sec:method} we present our method in detail and provide a primer in gravitational-wave parameter estimation.
We then apply our method to the first 11 compact binary merger gravitational-wave signals detected, presenting the results in Section~\ref{sec:results}. 
We conclude with a summary and discussion of some caveats to the method we have described.

\section{Marginalizing over PSD uncertainty} \label{sec:method}
The time-varying data in a gravitational-wave interferometer can be written in terms of an astrophysical signal, $h(\theta)$, and a noise term $n$:
\begin{align}
    d = h(\theta) + n.
\end{align}
For signals from \acp{CBC} with quasi-circular orbits, $\theta$ represents the 17 parameters describing the binary including the masses, tidal deformabilities and vector spins of the components, the sky location, the distance and inclination angle relative to the source, a polarization angle, and the time and phase at coalescence. 
The noise in each detector is typically assumed to be Gaussian and stationary~\cite{LIGOScientific:2019hgc}, such that the noise covariance matrix is diagonal in the frequency domain~\cite{Romano:2016dpx}:
\begin{align}
\langle \tilde{n}^{*}_{i}\tilde{n}_{j} \rangle = \frac{T}{2}S_{n}(f)\delta_{ij},
\end{align}
where $\tilde{n}$ denotes the Fourier transform of the noise contribution, the indices correspond to different frequency bins, $\delta_{ij}$ is the Kronecker delta, and $S_{n}(f)$ is the noise \ac{PSD} for that detector.

Under the assumption of stationary, Gaussian noise, the likelihood of observing data $d$ in one detector given the signal $h(\theta)$ and the power spectral density $S_{n}(f)$ is~\cite{Veitch:2014wba, Romano:2016dpx}: 
\begin{align}
    p(d | \theta, S_{n}) = \prod_{i} \frac{2}{\pi T S_{n}(f_{i})}\exp{\left[-\frac{2|\tilde{d}(f_{i}) - \tilde{h}(f_{i};\theta)|^{2}}{TS_{n}(f_{i})}\right]},
    \label{eq:likelihood}
\end{align}
where $T$ is the duration of the analyzed segment. 
When using data from multiple detectors, the joint likelihood is obtained by multiplying the individual likelihoods for each detector, while requiring the signal $h(\theta)$ to be coherent across the detector network:
\begin{align}
    p(\{d\}| \theta, \{S_{n}\}) = \prod_{j}^{N_{\mathrm{IFO}}} p(d_{j} | \theta, S_{n, j}),
\end{align}
where the index $j$ indicates the interferometer, $N_{\mathrm{IFO}}$ is the total number of interferometers in the network, and the signal parameters $\theta$ are assumed to be the same in all detectors. 
As mentioned in the previous section, the \ac{PSD} is usually assumed to be measured with infinite precision and is typically computed in one of two ways, although we will focus on the on-source method for the rest of this paper.

This method is implemented in the \bw package, which uses the \bl algorithm to fit the \ac{PSD}~\cite{Littenberg:2014oda,Cornish:2014kda}. 
This algorithm uses the same likelihood defined in Eq.~\ref{eq:likelihood}, with the exception that the astrophysical contribution $h$ is no longer a waveform that depends on the 17 binary parameters $\theta$ but rather a sum of wavelets. 
When \bw is used to characterize only the properties of the noise without assuming the presence of an astrophysical signal, it is run independently for each interferometer and includes a glitch model that allows for the presence of independent non-Gaussian noise excursions in each detector. 
These glitches are modeled separately from the background Gaussian noise using wavelets without the requirement that the reconstructed glitch signal be coherent across the different detectors\footnote{We note that in this configuration, the ``glitch model'' is also expected to capture any \ac{GW} signal present, such that the noise model only describes the Gaussian contribution to the data}.
\bw samples over the properties of the background Gaussian noise model, again characterized by a variable number of spline points and Lorenztians, recording the parameters of those components as a posterior sample.
These components can also be represented as an instance of a posterior set of \acp{PSD} describing the inferred variance of the Gaussian noise in the analyzed data.
Recent \ac{PE} analyses of \ac{CBC} \ac{GW} events~\cite{LIGOScientific:2018mvr} have only considered describing the \ac{PSD}, the $S_{n}$ term in the likelihood of Eq.~\ref{eq:likelihood}, through a fixed point estimate of the overall \ac{PSD} posterior distribution inferred by \bw.
As shown by~\cite{Chatziioannou:2019zvs}, the preferred point estimate in that case is the median \ac{PSD}.
It is obtained by evaluating the median value of $S_{n}(f)$ in each frequency bin from among the individual PSDs computed for each of the posterior samples \bw inferred for the splines and Lorentzians.
This means that the median \ac{PSD} by construction does not correspond to any individual \ac{PSD} posterior sample, and that it formally is not required to be smooth over adjacent bins, something that is enforced by construction by the spline instance in each posterior \ac{PSD}.

Ideally, the binary parameters $\theta$ and the \ac{PSD} would be estimated simultaneously using the likelihood above, but this is at present prohibitively difficult since fitting the spline and Lorentzian parameters describing the \ac{PSD} requires a trans-dimensional \ac{MCMC} algorithm~\cite{Littenberg:2014oda}, and current frameworks for \ac{CBC} parameter estimation depend on using fixed-dimensional models~\cite{Veitch:2014wba, Ashton:2018jfp}. %
We thus write the combined posterior for the \ac{PSD} and the binary parameters as the product of two separate posterior probabilities, under the assumption that the binary signal parameters and the \ac{PSD} are uncorrelated:
\begin{align}
p(\theta, S_{n} | d) = p(\theta | S_{n}, d) p(S_{n} | d) 
\label{eq:joint_post}
\end{align}

To obtain the posterior on $\theta$ we marginalize the above expression over the \ac{PSD}:
\begin{align}
    p(\theta|d) =   \int{\mathrm{d}S_{n} \; p(\theta, S_{n} | d)} = \int{\mathrm{d}S_{n} \; p(\theta |S_{n}, d) p(S_{n}|d)},
    \label{eq:marg_post_int}
\end{align}
which is the expectation value of the posterior on the binary parameters averaged over the \ac{PSD} uncertainty.

In practice, we first obtain a discrete set of $N$ posterior samples for the \ac{PSD} using \bw.
For each \ac{PSD} posterior sample $k$, we run Bayesian \ac{PE} and obtain a posterior on the binary parameters via:
\begin{align}
    p(\theta | S_{n,k}, d) = \frac{\pi(\theta)}{\mathcal{Z}_{k}}p(d |\theta, S_{n, k}), \label{eq:individual_post}
\end{align}
where $\pi(\theta)$ is the prior, and the denominator is the evidence, or marginalized likelihood, for a particular PSD posterior sample:
\begin{align}
    \mathcal{Z}_{k} = p(d|S_{n,k}) = \int p(d |\theta, S_{n, k})\pi(\theta) \mathrm{d}\theta.
    \label{eq:evidence}
\end{align} 
Because Eq.~\ref{eq:marg_post_int} is just the expectation value of the binary posteriors obtained with each of the \ac{PSD} posterior samples, it can be rewritten as a sum of the individual posteriors:
\begin{align}
    p(\theta | d) = \frac{1}{N}\sum_{k} p(\theta |S_{n, k}, d).
    \label{eq:marginalized_post}
\end{align}

\section{Application to current gravitational wave detections} \label{sec:results}
We apply the method described in the previous section to each of the 10 \acp{BBH} in the first gravitational wave transient catalog, along with the \ac{BNS} merger, GW170817~\cite{LIGOScientific:2018mvr, TheLIGOScientific:2017qsa}. 
For the \acp{BBH} we first run \bw to generate 200 fair draws from the \ac{PSD} posterior for each event, while for the \ac{BNS} we only use 141 fair draws to restrict the computational cost. 
We discuss the choice of the number of \ac{PSD} posterior samples used for marginalization further in Section~\ref{sec:discussion}. To contain the computational cost, we pair the first \ac{PSD} posterior sample for one detector with the first for the other detectors, so we only generate 200 and 141 total \ac{PSD} posterior sample pairs for \ac{BBH} and \ac{BNS} respectively. 
For each of the \ac{PSD} sample pairs, we use the \bilby \ac{PE} package~\cite{Ashton:2018jfp,Bilby-code} with the \texttt{dynesty} nested sampler~\cite{Dynesty} to obtain a posterior distribution for $\theta$, the 17 binary parameters\footnote{For assumed \ac{BBH} sources, we fix the tidal deformability parameters to 0.}.
For \ac{BBH} sources, we use the \texttt{IMRPhenomPv2} waveform~\cite{Husa:2015iqa,Khan:2015jqa,Hannam:2013oca}, while for the \ac{BNS} we use the \texttt{IMRPhenomPv2\_NRTidal} waveform~\cite{Dietrich:2017aum, Dietrich:2018uni} implemented via a reduced order quadrature likelihood~\cite{Smith:2016qas} to reduce the computational cost.

We use priors that are uniform in the chirp mass,
\begin{align}
    \mathcal{M} \equiv \frac{(m_1 m_2)^{3/5}}{(m_1+m_2)^{1/5}},
\end{align} 
asymmetric mass ratio,
\begin{align}
    q\equiv m_2/m_1\;\mathrm{ with }\; m_2\leq m_1
\end{align} 
and dimensionless spin magnitudes, and proportional to the square of the luminosity distance, $d_{L}$. 
The mass ratio prior ranges from $q=1/8$ to 1 for all events except GW151012, where the lower bound of the mass ratio prior was extended to $1/17.95$ because of posterior support at lower mass ratios. 
The spin magnitude prior covers the range $a \in [0,0.99]$ for \acp{BBH}, while for the \ac{BNS} we use a restricted spin prior covering the range $a \in [0,0.05]$ motivated by the component spins of observed galactic double neutron star systems~\cite{Burgay:2003jj, TheLIGOScientific:2017qsa}. 
The priors on the tidal parameters for the \ac{BNS}, $\Lambda_{1}$ and $\Lambda_{2}$, are independent and uniform from 0 to 5000.
The marginalized posteriors are obtained by combining the samples from all 200 runs, choosing 5000 samples from each run since they all have equal weights according to Eq.~\ref{eq:marginalized_post}, and the result files are available for download on git~\cite{data_release}.
We also perform an analysis with the median \ac{PSD} computed by \bw in order to emulate the analyses typically performed by LIGO/Virgo and to compare with the marginalized posteriors. 
The run settings are described in Table~\ref{tab:settings}.
\begin{table*}[htb]
    \centering
\begin{tabular}{|p{2cm} ||p{2cm} p{2cm} p{2cm} p{2cm} p{2cm} p{2cm}|}
    \hline
     Event & Duration [s] & $f_{max}$ [Hz] & $\mathcal{M}_{\min}~\mathrm{[M_{\odot}]}$ &$\mathcal{M}_{max}~\mathrm{[M_{\odot}]}$ & $d_{L, \min}~[\mathrm{Mpc}]$ & $d_{L,\max}~[\mathrm{Mpc}]$ \\
    \hline \hline
    GW150914 & 4 & 1024 & 9 & 69.9 & 500 & 2000\\
    GW151012 & 4 & 512 & 10 & 30 & 223 & 3000\\
    GW151226 & 8 & 512 & 5 & 12.3 & 20 & 1500\\
    GW170104 & 4 & 1024 & 12.3 & 45 & 40 & 3260\\
    GW170608 & 16 & 1024 & 5 & 12.3 & 90 & 2000\\
    GW170729 & 4 & 1024 & 9 & 173 & 226 & 7000\\
    GW170809 & 4 & 1024 & 12.3 & 43.5 & 168 & 4000\\
    GW170814 & 4 & 1024 & 12.3 & 45 & 217 & 3000\\
    GW170817 & 128 & 2048 & 1.18 & 1.21 & 1 & 75\\
    GW170818 & 4 & 1024 & 12.3 & 45 & 100 & 2000\\
    GW170823 & 4 & 1024 & 12.3 & 55 & 286 & 6000\\
    \hline
\end{tabular}
\caption{Run settings for each compact binary signal analyzed. The prior limits on the chirp mass are specified in the detector frame.
The starting frequency for the overlap integral is 20 Hz for all \ac{BBH} events with the exception of GW170608, which suffered from low-frequency noise in the Hanford interferometer, so the starting frequency was chosen to be 30 Hz~\cite{Abbott:2017gyy}. 
The starting frequency for the \ac{BNS} analysis was 23~Hz, in accordance with the range of validity of the reduced order quandrature used.}
\label{tab:settings}
\end{table*}

\begin{figure*}
\includegraphics[width=\textwidth]{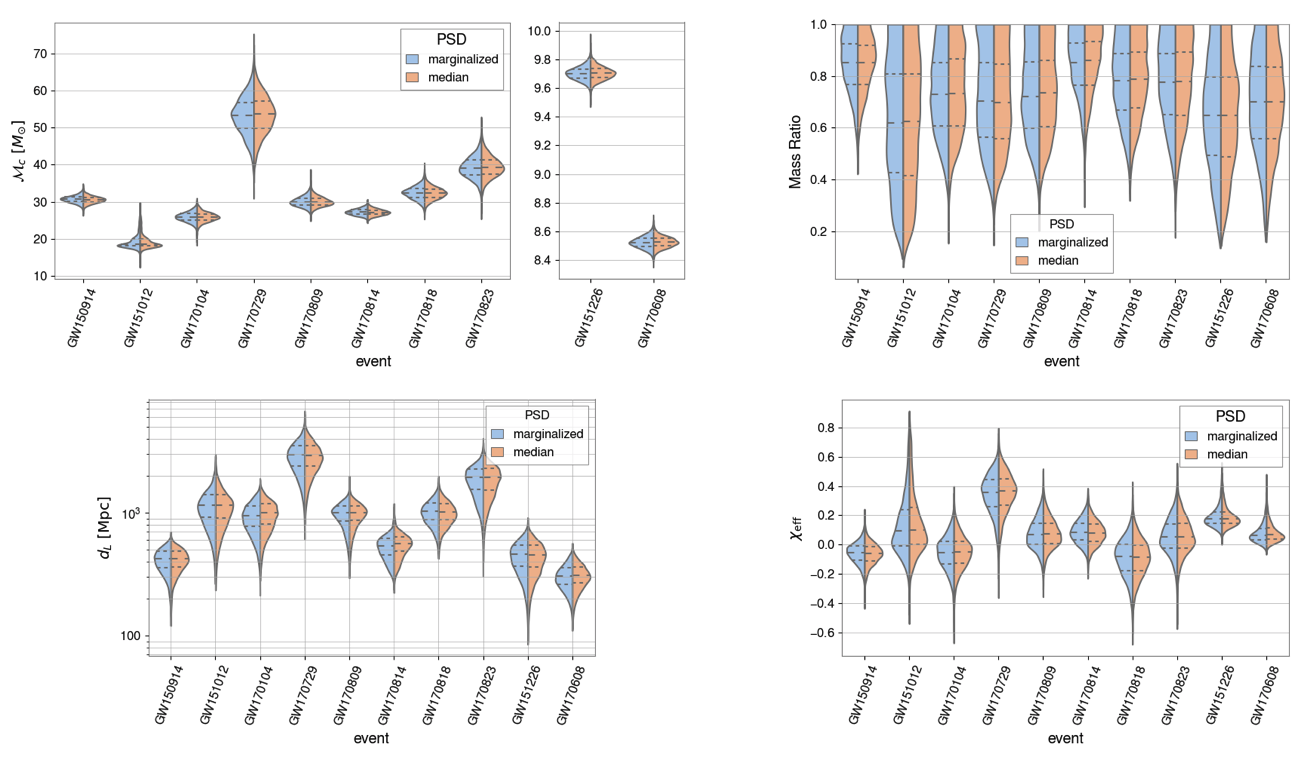}
\caption{Violin plots showing the probability density for the \ac{PSD}-marginalized posterior samples (blue) compared to the posterior samples obtained using the median \ac{PSD} (orange) for the chirp mass, mass ratio, luminosity distance, and effective spin for each of the \ac{BBH} detections.
The horizontal lines represent the median (dashed) and $1\sigma$ confidence intervals (dotted) for each event.
}
\label{fig:violins}
\end{figure*}

\begin{figure}[!htbp]
\includegraphics[width=0.47\textwidth]{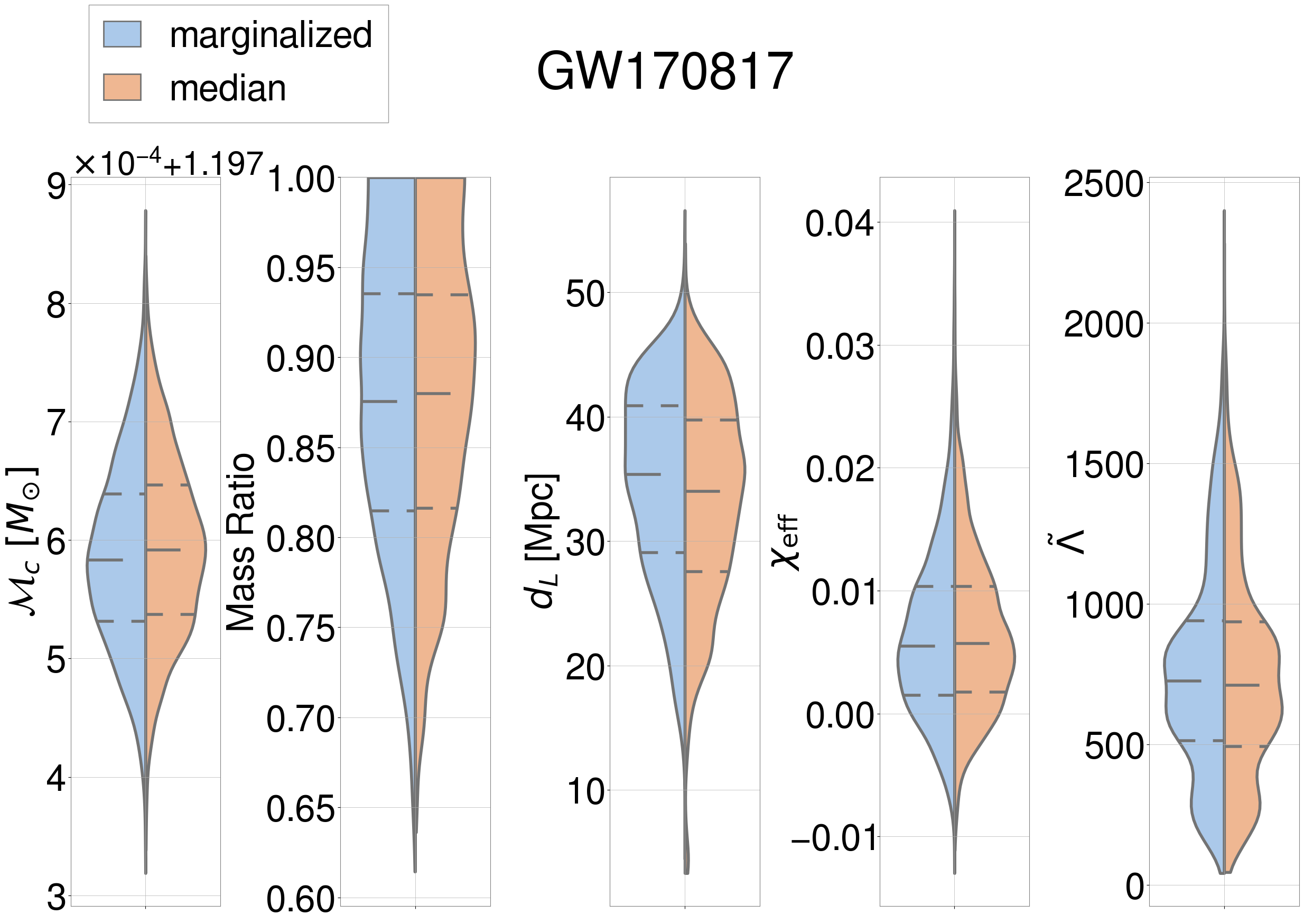}
\caption{Violin plots showing the probability density for the \ac{PSD}-marginalized posterior samples (blue) compared to the posterior samples obtained using the median \ac{PSD} (orange) for the chirp mass, mass ratio, luminosity distance, effective spin, and mass-weighted average tidal deformability for the \ac{BNS}, GW170817.
The horizontal lines represent the median (dashed) and $1\sigma$ confidence intervals (dotted) for each parameter.}
\label{fig:170817_violins}
\end{figure}

Fig.~\ref{fig:violins} shows the \ac{PSD}-marginalized posteriors for the detector frame chirp mass, mass ratio, effective spin $(\chi_{\mathrm{eff}})$, and luminosity distance compared to the posteriors obtained using the median \ac{PSD} as a point estimate for the detected \acp{BBH}, and Fig.~\ref{fig:170817_violins} shows the same results for the \ac{BNS}, in addition to the comparison of the mass-weighted average tidal deformability, $\tilde{\Lambda}$.
The effective spin is the mass-weighted projection of the component spins along the direction of the orbital angular momentum~\cite{Campanelli:2006uy, Racine:2008qv} and is the best measured spin parameter with gravitational wave data~\cite{Vitale:2016avz, Ng:2018neg}. 
$\tilde{\Lambda}$ determines the gravitational-wave phase to leading order in the individual tidal deformabilities, $\Lambda_{1}$ and $\Lambda_{2}$, and is defined as~\cite{Flanagan:2007ix, Favata:2013rwa}:
\begin{align}
    \tilde{\Lambda} = \frac{16}{13}\frac{(m_{1} + 12m_{2})m_{1}^{4}\Lambda_{1} + (m_{2}+12m_{1})m_{2}^{4}\Lambda_{2}}{(m_1+m_2)^{5}}.
\end{align}
While small variations in posterior shape and position are observed, marginalizing over the \ac{PSD} uncertainty generally appears to lead to only a minor increase in the width of the binary parameter posteriors. 

\begin{figure}[!htbp]
\includegraphics[width=0.5\textwidth]{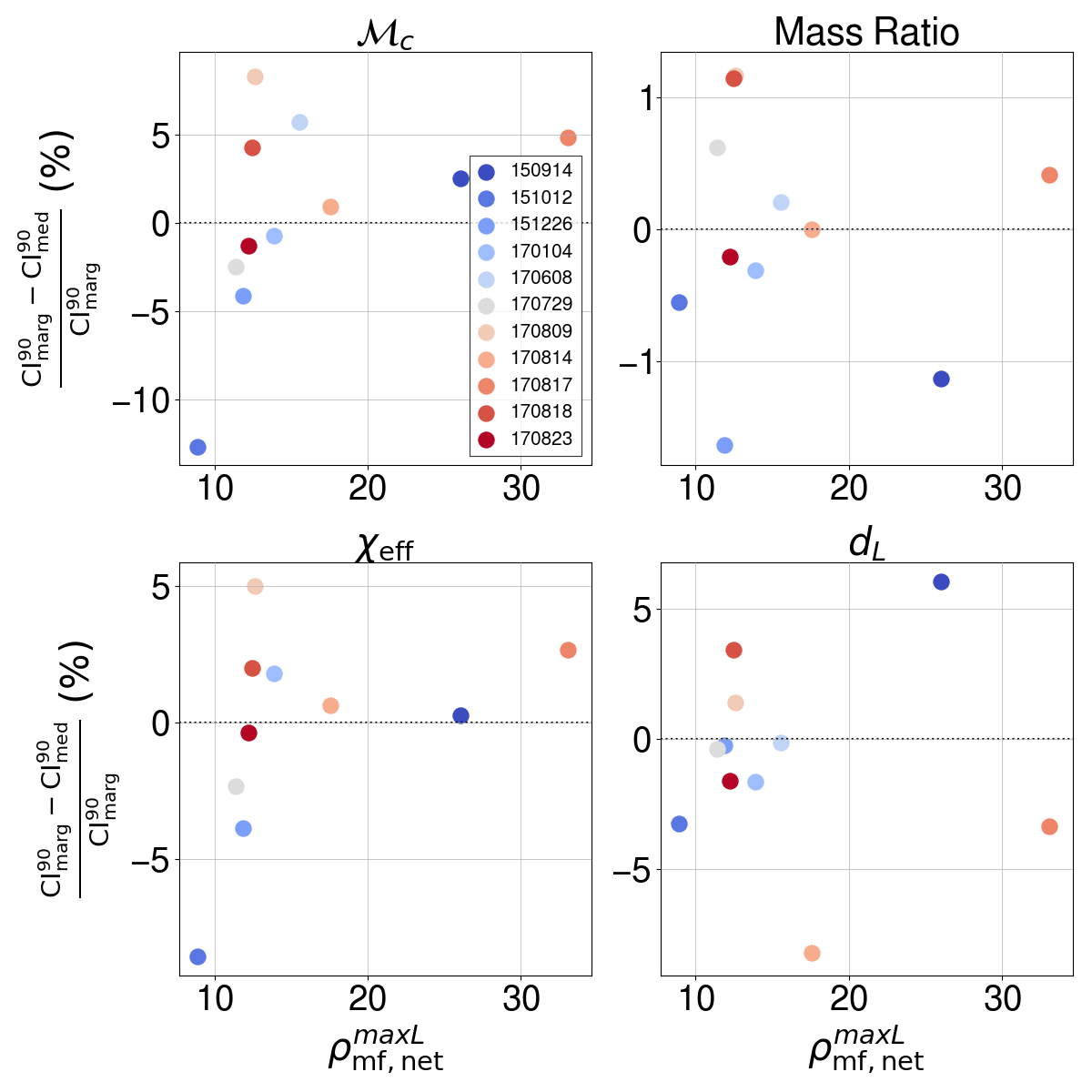}
\caption{Normalized difference between the width of the 90\% confidence interval of the \ac{PSD}-marginalized and median \ac{PSD} posteriors for chirp mass, mass ratio, effective spin, and luminosity distance for each event as a function of the maximum-likelihood network matched filter \ac{SNR} calculated using the median \ac{PSD}.}
\label{fig:delta_vs_snr}
\end{figure}

\begin{figure}[!htbp]
\includegraphics[width=0.5\textwidth]{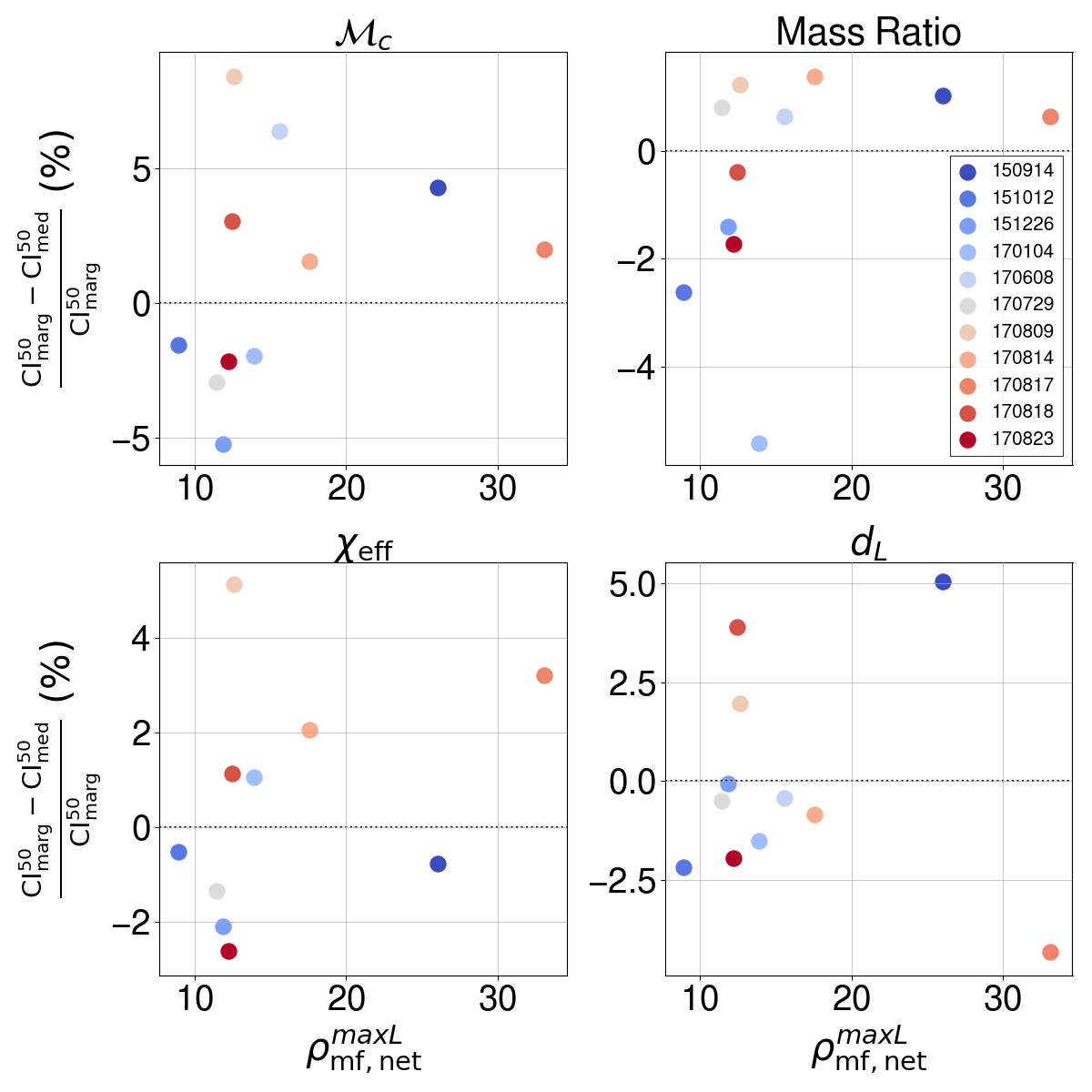}
\caption{Normalized difference between the width of the 50\% confidence interval of the \ac{PSD}-marginalized and median \ac{PSD} posteriors for chirp mass, mass ratio, effective spin, and luminosity distance for each event as a function of the maximum-likelihood network matched filter \ac{SNR} calculated using the median \ac{PSD}.}
\label{fig:delta_vs_snr_50}
\end{figure}

To quantify this effect, we show the difference in the width of the 90\% and 50\% confidence intervals between the \ac{PSD}-marginalized and median \ac{PSD} posteriors in Figs.~\ref{fig:delta_vs_snr} and \ref{fig:delta_vs_snr_50} respectively for each event as a function of the network matched filter \ac{SNR} calculated at the maximum likelihood point. 
The network \ac{SNR} is obtained by adding the individual-detector \acp{SNR} in quadrature, where the matched filter \ac{SNR} in a single detector is given by\cite{Allen:2005fk}:
\begin{align}
    \rho_{mf} = \frac{\langle d, h\rangle}{\sqrt{\langle h, h \rangle}},
\end{align}
for the inner product defined as:
\begin{align}
    \langle a, b\rangle = 4\Re\int_{0}^{\infty}\frac{\tilde{a}^{*}(f)\tilde{b}(f)}{S_{n}(f)}\mathrm{d}f.
\end{align}
These results indicate that marginalizing over the uncertainty in the \ac{PSD} produces posteriors that are wider than those obtained with the median \ac{PSD} as a point estimate for about half of the events. The fractional change of the posterior width for both the 90\% and 50\% confidence intervals is of the order of a few percent, although a few larger excursions are observed for both confidence intervals. No significant trend is observed in the change in the confidence interval width as a function of the \ac{SNR}, which indicates that the properties of the noise and the subsequent variation in the \ac{PSD} are independent of the strength of the signal. 

\begin{figure}[!htbp]
\includegraphics[width=0.5\textwidth]{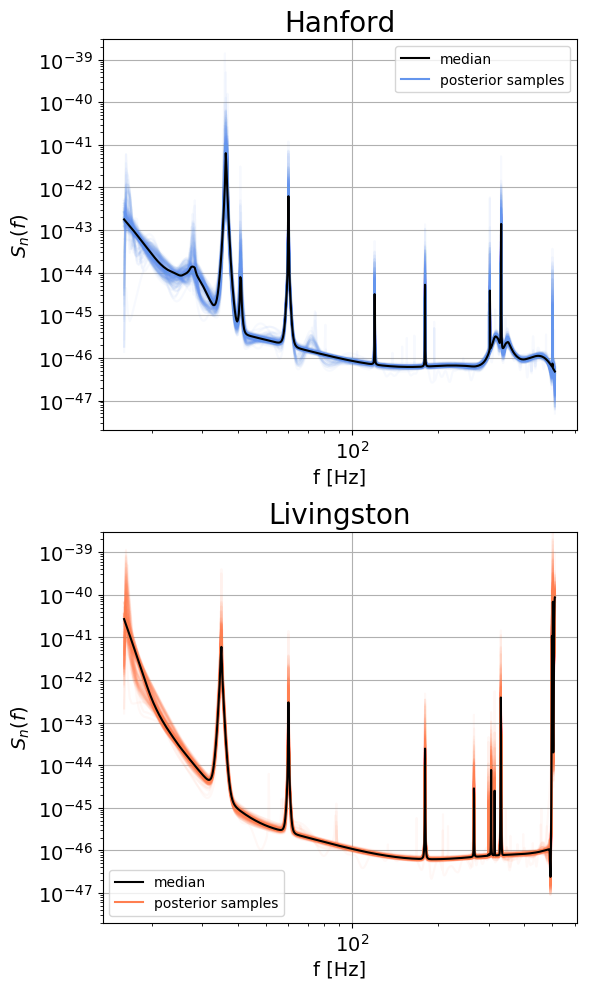}
\caption{The 200 analyzed \ac{PSD} posterior samples for the Hanford and Livingston LIGO detectors for GW151012 along with the median \ac{PSD} in black}
\label{fig:psds_plotted}
\end{figure}

\begin{figure}[!htbp]
\includegraphics[width=0.5\textwidth]{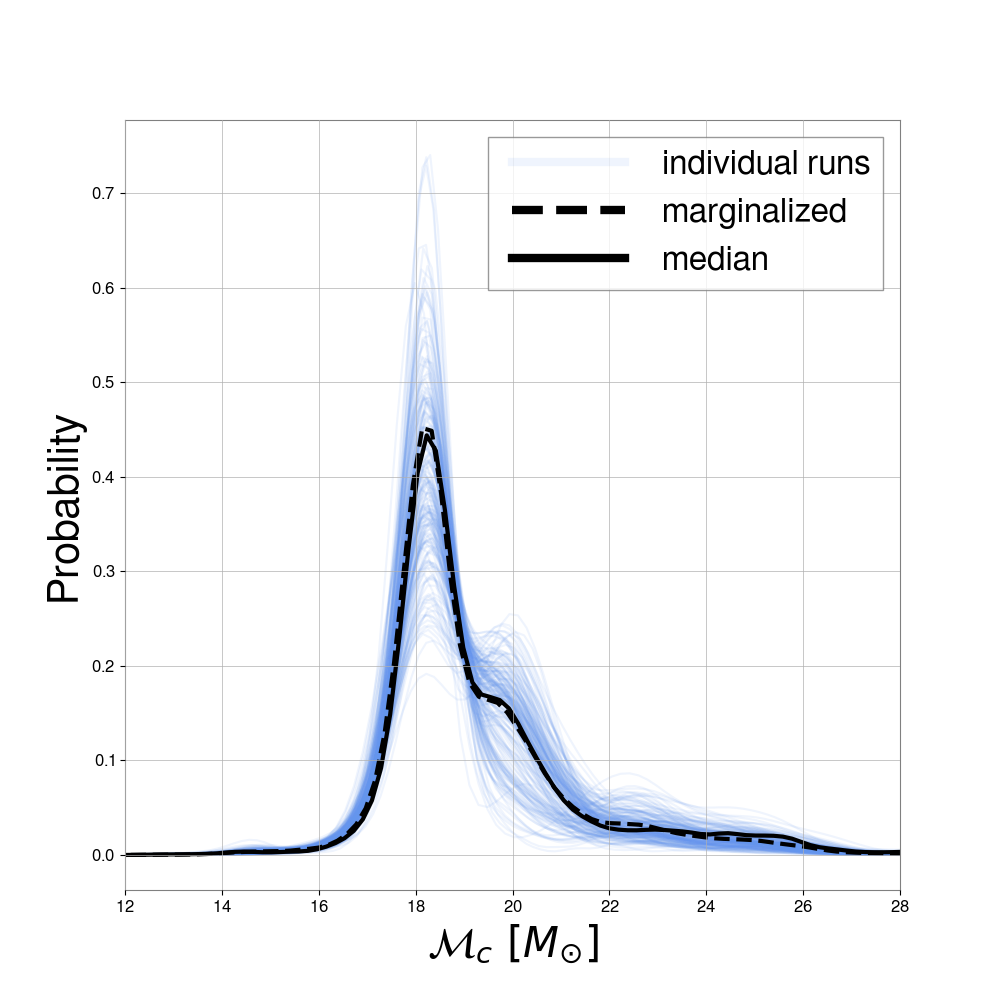}
\caption{Kernel density estimate of the posterior on the detector-frame chirp mass for all 200 individual \ac{PSD} posterior samples with the total \ac{PSD}-marginalized posterior shown in the dotted black line and the posterior obtained with the median \ac{PSD} in the solid black line for GW151012.}
\label{fig:151012_chirp_mass}
\end{figure}

The largest deviation occurs in the 90\% confidence interval of the chirp mass posterior of GW151012. This behavior can be explained by the variability in the Hanford \ac{PSD} posterior at low frequencies shown in Fig.~\ref{fig:psds_plotted} that the median \ac{PSD} cannot account for. This variability translates into significant posterior support for higher chirp masses for some of the individual \ac{PSD} posterior samples, as shown in Fig.~\ref{fig:151012_chirp_mass}. This effect is minimized when averaging over the full \ac{PSD} posterior, but not for the run with the median \ac{PSD} alone, which also has increased support for higher chirp masses. Because the posterior obtained with the median \ac{PSD} has wider tails, it has a correspondingly wider 90\% confidence interval without affecting the 50\% confidence interval, indicating good agreement between the two results for the bulk of their posterior distributions.


Table~\ref{tab:sky_area} shows the fractional change in the sky area $\Omega$, in square degrees, contained within the 50\% and 90\% confidence intervals between the \ac{PSD}-marginalized and median \ac{PSD} posteriors, $\Delta \Omega = (\Omega_{\mathrm{marg}} - \Omega_{\mathrm{med}})/\Omega_{\mathrm{marg}}$, as well as the absolute difference in the sky area contained in the 90\% confidence interval. 
The change in the confidence intervals for the sky area exhibits similar variation to the other parameters shown in Figs.~\ref{fig:delta_vs_snr} and \ref{fig:delta_vs_snr_50}, on the order of $\sim 10\%$. The biggest deviations occur for the best-localized event, GW170817, although the total change is only a few square degrees for both confidence intervals. 

\begin{table}
    \centering
\begin{tabular}{|p{2cm} ||p{1.5cm} p{1.5cm} p{2cm}|}
    \hline
     Event & $\Delta \Omega_{50}~(\%)$ & $\Delta \Omega_{90}~(\%)$ & $\Delta \Omega_{90}~(\mathrm{deg}^{2})$ \\
    \hline \hline
    GW150914 & 13.7 & 12.3 & 21 \\
    GW151012 & 1.5 & -1.1 & -20 \\
    GW151226 & -9.6 & -6.7 & -93 \\
    GW170104 & 15.3 & 7.5 & 77 \\
    GW170608 & -12.0 & 1.9 & 8 \\
    GW170729 & 19.0 & 10.4 & 136\\
    GW170809 & -12.1 & 2.9 & 9 \\
    GW170814 & 0 & -18.6 & -24 \\
    GW170817 & 28.6 & 25.9 & 7 \\
    GW170818 & 11.1 & 6.5 & 2 \\
    GW170823 & 2.9 & -0.9 & -14 \\
    \hline
\end{tabular}
\caption{Fractional change in the 50\% and 90\% confidence intervals for the sky area in square degrees between the \ac{PSD}-marginalized and median-\ac{PSD} posteriors for each event, defined as $\Delta \Omega = (\Omega_{\mathrm{marg}} - \Omega_{\mathrm{med}})/\Omega_{\mathrm{marg}}$, and the absolute change in square degrees for the 90\% confidence interval, which is $\Omega_{\mathrm{marg}} - \Omega_{\mathrm{med}}$.}
\label{tab:sky_area}
\end{table}

In order to compare the posterior variations due to marginalizing over the \ac{PSD} uncertainty to those expected due to statistical fluctuations, we use bootstrapping to generate different sets of samples from the median \ac{PSD} posterior for different parameters and events. We find that on average the change in the width of the 90\% confidence interval among the bootstrapped samples is of the order of $\sim 0.1\%$ across different events and parameters, about an order of magnitude smaller than the deviations we observe due to marginalizing over the \ac{PSD} uncertainty.

\section{Discussion} \label{sec:discussion}
In this paper, we demonstrate a new method for marginalizing over the uncertainty in the noise power spectral density when performing gravitational wave parameter estimation for compact binary sources. 
We first obtain posterior samples for the \ac{PSD} itself using the \bw software and then perform parameter estimation using 200 fair draws from the \ac{PSD} posterior, combining the samples with equal weights to obtain the \ac{PSD}-marginalized posterior. 
While no critical difference is observed in the posterior peak and shape for the binary parameters between the posteriors obtained using the median \ac{PSD} as a point estimate and the \ac{PSD}-marginalized posteriors, the posterior widths including the sky area vary on the order of a few percent, with the \ac{PSD}-marginalized posterior being broader than that obtained using the median \ac{PSD} for about half of the events.
We do find more significant variation in the binary parameter posteriors obtained using individual \ac{PSD} posterior samples for each gravitational-wave event, which can pick up secondary peaks or stronger support in the tail of the posterior, as was the case for GW151012. 
Based on these results, we conclude that the median \ac{PSD} provides posteriors whose position and width are similar to those obtained when marginalizing over the PSD uncertainty to within a few percent for gravitational-wave signals of these SNRs observed in detectors with the current noise properties. The variations between the median and \ac{PSD}-marginalized posteriors are an order of magnitude larger than those expected due to statistical fluctuations.

We close with a discussion of some caveats to our analysis. While we have shown that there is only minimal variation between the binary parameter posteriors obtained by marginalizing over the uncertainty in the \ac{PSD} measurement and those obtained using the median PSD as a point estimate, we have not determined which of the two \ac{PSD} models is preferred by the data. In our case, the two models we wish to compare are the ``varied spline and Lorentzian model'', where the \ac{PSD} is allowed to vary and is parameterized in terms of a broadband spline and a series of Lorentzians to fit narrowband features, and the ``median \ac{PSD} model'', where the \ac{PSD} is fixed to the median of the full posterior computed with \bw. In both cases, the model also includes the presence of a \ac{CBC} signal in the data, so we denote the varied spline and Lorentzian model as CBC+SL and the median \ac{PSD} model as CBC+Me. This question of which model is statistically preferred is typically answered via Bayesian model selection and quantified using a Bayes factor between the two models being compared:
\begin{align}
    \mathrm{BF}^{\mathrm{CBC+SL}}_{\mathrm{CBC+Me}} = \frac{\mathcal{Z}^{\mathrm{CBC+SL}}}{\mathcal{Z}^{\mathrm{CBC+Me}}},
\end{align}
where the evidence, $\mathcal{Z}$, is defined as the normalization factor of the posterior obtained using a particular model. In addition to comparing the CBC+SL and CBC+Me models, obtaining the evidence under the CBC+SL model could also be used for comparing different variations of the \ac{CBC}+SL model, for example precessing versus aligned compact binary component spins, the presence of higher order modes in the \ac{CBC} waveform, or deviations from general relativity, which would all be represented as $\mathcal{Z}^{\mathrm{CBC+X+SL}}$.

Using the method described in Eq.~\ref{eq:marginalized_post}, we obtain posterior samples for for the CBC+SL model but not an evidence:
\begin{align}
\mathcal{Z}^{\mathrm{CBC+SL}} = p(d|\mathrm{CBC+SL}) &= \int p(d | \theta, \mathrm{CBC+SL}) \\
        &\times \pi(\theta |\mathrm{CBC+SL})\mathrm{d}\theta, \nonumber
\end{align}
where $\pi(\theta |\mathrm{CBC+SL})$ is the prior defined in Eq.~\ref{eq:individual_post} where the explicit dependence on the $\mathrm{CBC+SL}$ model had previously been suppressed. $p(d|\mathrm{CBC+SL})$ is the likelihood of the data given the $\mathrm{CBC+SL}$ model, which we need to define in order to calculate the evidence above.

The evidence can be calculated during sampling if instead of marginalizing over the \ac{PSD} uncertainty by combining the binary parameter posteriors obtained using different \ac{PSD} posterior samples, the likelihood is modified to account for the \ac{PSD} uncertainty, and this modified likelihood is used to estimate the \ac{PSD}-marginalized binary parameters directly. 
The marginalized likelihood for the $\mathrm{CBC+SL}$ model is given by:
\begin{align}
    p(d|\theta, \mathrm{CBC+SL}) &= \int p(d|\theta,S_{n}, \mathrm{CBC+SL})\\
    &\times \pi(S_{n}|\mathrm{CBC+SL})\mathrm{d}S_{n}. \nonumber
\end{align}
However, the likelihood $p(d|\theta,S_{n}, \mathrm{CBC+SL})$ doesn't depend on whether the PSD uncertainty is being included; it is equivalent to the likelihood for CBC signals defined in Eq.~\ref{eq:likelihood}, so we drop the explicit dependence on $\mathrm{SL}$. 
Similarly, the prior on the \ac{PSD}, $\pi(S_{n}|\mathrm{CBC+SL})$ doesn't depend on the presence of a CBC signal, so we drop the dependence on $\mathrm{CBC}$:
\begin{align}
    \label{eq:marg_like}
    p(d|\theta, \mathrm{CBC+SL}) &= \int p(d|\theta,S_{n}, \mathrm{CBC})\\ 
    &\times \pi(S_{n}|\mathrm{SL})\mathrm{d}S_{n}. \nonumber
\end{align}

The prior on the PSD under the varied spline and Lorentzian model is the prior used by \bw, which is actually a complicated function of the spline and Lorentzian parameters and cannot be straightforwardly expressed in terms of $S_{n}$ directly. Unfortunately, it also cannot be extracted from the \bw sampler products. Since \bw must be run in the configuration where it also models the non-Gaussian data component from the astrophysical \ac{CBC} signal in order to obtain an unbiased estimate of the \ac{PSD}, the prior that can be constructed from the sampler products is a joint prior on the spline and Lorentzian parameters and the wavelet parameters used to model the non-Gaussian component. Thus, the marginalized likelihood in Eq.\ref{eq:marg_like} cannot be obtained using the two-step system we have employed in the rest of the analysis where the \ac{PSD} is estimated first, separately from the estimation of the binary parameters. One possible way to obtain the evidence for the CBC+SL model would be to simultaneously estimate both the binary parameters and the \ac{PSD} using the spline and Lorentzian parameterization. 

We note that the likelihood under the CBC+Me model can be recovered from the form of the likelihood in Eq.~\ref{eq:marg_like} by substituting a Dirac delta for the prior on $S_{n}$: 
\begin{align}
    \pi(S_{n}|\mathrm{SL})\rightarrow{} \pi(S_{n} | \mathrm{Me}) = \delta(S_{n} - S_{n,\mathrm{Me}}).
\end{align}
The evidence for this model is hence obtained during the sampling of the binary parameters using the median \ac{PSD}.

While our method is embarrassingly parallel, so that each of the $N$ parameter estimation analyses with different \acp{PSD} can be launched simultaneously, using the modified likelihood above requires that the CBC signal likelihood in Eq.~\ref{eq:likelihood} is evaluated $N$ times in series for each binary parameter sample. 
This computation could in principle be accelerated through the use of likelihood reweighting~\cite{Payne:2019wmy} or other parallel processing techniques such as graphical processing units (GPUs)~\cite{Talbot:2019okv}, multiprocessing~\cite{Smith:2019ucc}, or the Message Passing Interface (MPI)~\cite{mpi}.
Furthermore, our method still requires $N$ times the computational resources compared to using the median \ac{PSD} as a point estimate. 
We have chosen $N$=200 somewhat arbitrarily, balancing the computational cost against the number of samples needed to adequately represent the complete \ac{PSD} posterior. 
We note, however, that producing the 200 draws from the \ac{PSD} posterior comes at no extra computational cost compared to the production of the median \ac{PSD}, so it would be possible to inspect the variability of the \ac{PSD} posterior before starting the follow-up parameter estimation with each posterior draw. 
While we did find in the case of GW151012 that increased variability in the \ac{PSD} posterior led to a larger change in the width of the binary parameter posteriors, there were also changes on the order of $\sim5\%$ for events whose \ac{PSD} posteriors seemed ``smooth" like for the mass ratio of GW170104.
We leave the systematic determination of the optimal $N$ to future work, along with the investigation of the applicability of the likelihood reweighting method to marginalization over \ac{PSD} uncertainty.

We emphasize that while \bw can account for the presence of non-Gaussian ``glitches'' in the data such that they do not affect the inference of the \ac{PSD} parameters, it still assumes that the noise is stationary and Gaussian over the duration of the analysis segment. 
Thus, the method we have described here does not include marginalizing over \ac{PSD} uncertainty due to the variation of the noise properties over time. 
This should not have a significant effect for short signals like \acp{BBH}, but can be more pronounced for \ac{BNS} signals requiring longer analysis segments~\cite{Chatziioannou:2019zvs, Zackay:2019kkv}.

\begin{acknowledgments}
S.B., S.V., and C.-J.H.~acknowledge support of the National Science Foundation, and the LIGO Laboratory. 
LIGO was constructed by the California Institute of Technology and Massachusetts Institute of Technology with funding from the National Science Foundation and operates under cooperative agreement PHY-1764464.
S.B. is also supported by the Paul and Daisy Soros Fellowship for New Americans and the NSF Graduate Research Fellowship under Grant No. DGE-1122374.
J.D was supported by Imperial College's International Research Opportunities Programme.
The authors would like to thank Katerina Chatziioannou, Will Farr, Max Isi, and Colm Talbot for insightful discussions, and Tyson Littenberg for helpful comments on the manuscript.
The authors are grateful for computational resources provided by the LIGO Laboratory and supported by National Science Foundation Grants PHY-0757058 and PHY-0823459.
This article carries LIGO Document Number LIGO-P2000123.
\end{acknowledgments}

\appendix
\section{PSD Posteriors}
\begin{figure*}
\includegraphics[width=\textwidth]{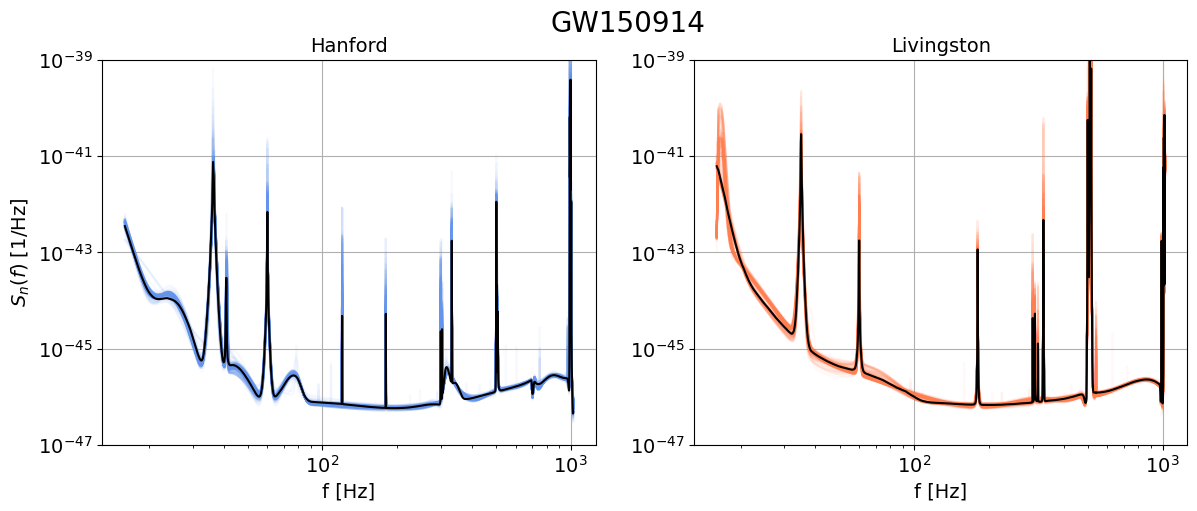}
\includegraphics[width=\textwidth]{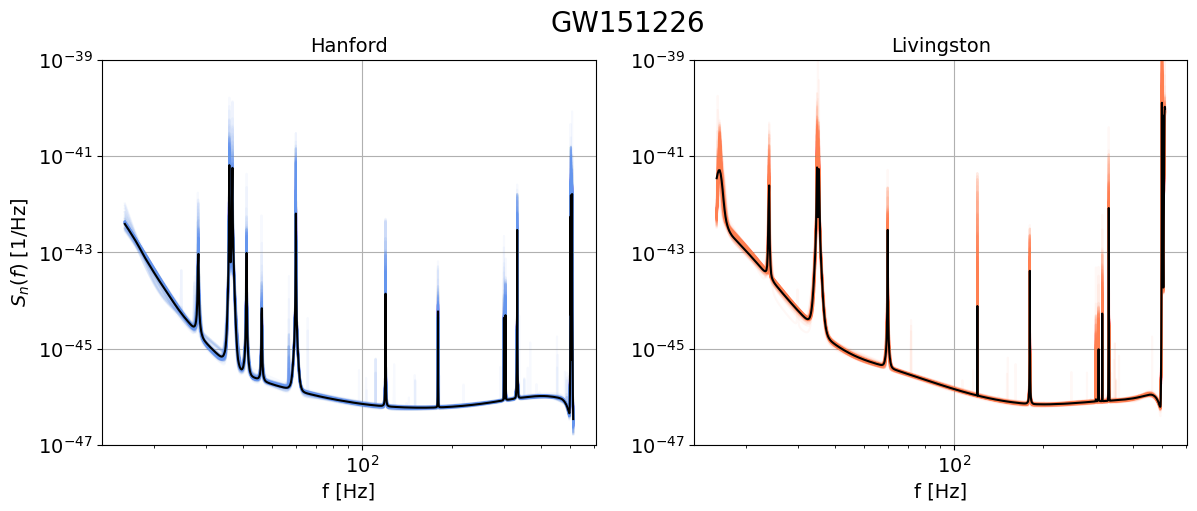}
\includegraphics[width=\textwidth]{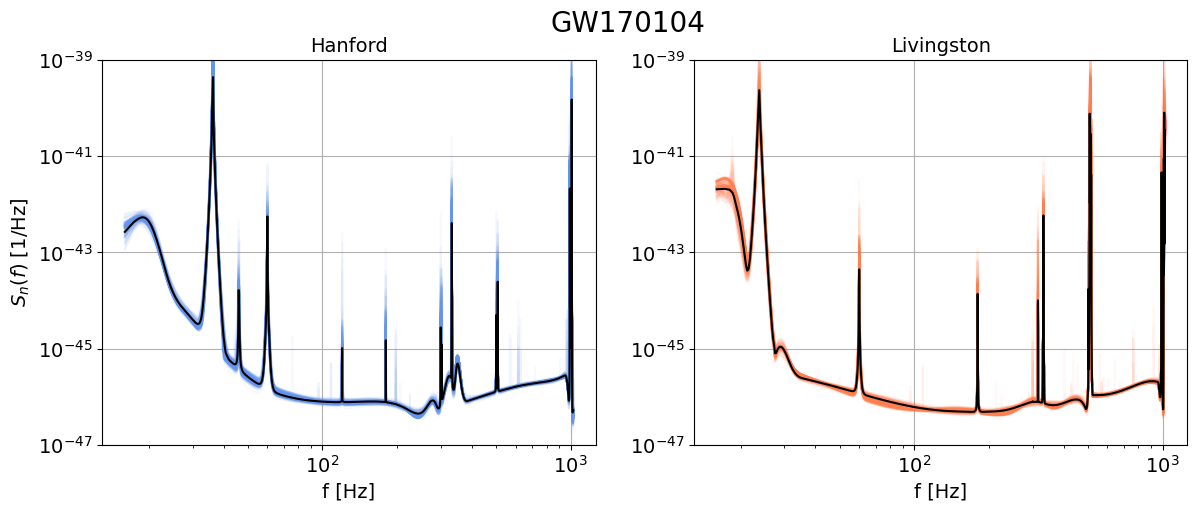}
\caption{\ac{PSD} posterior samples (colored lines) and the median \ac{PSD} (black line) for each of the 11 events in GWTC-1. The \acp{PSD} start at 16~Hz, and the maximum frequency for each event is given in Table.~\ref{tab:settings}.}
\label{fig:psds1}
\end{figure*}

\begin{figure*}
\includegraphics[width=0.66\textwidth]{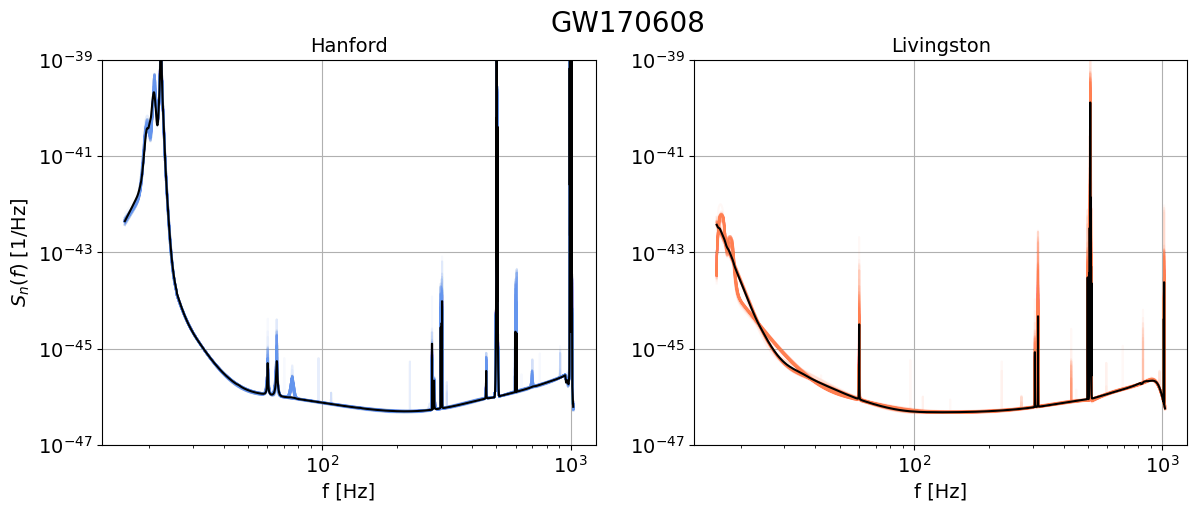}
\includegraphics[width=\textwidth]{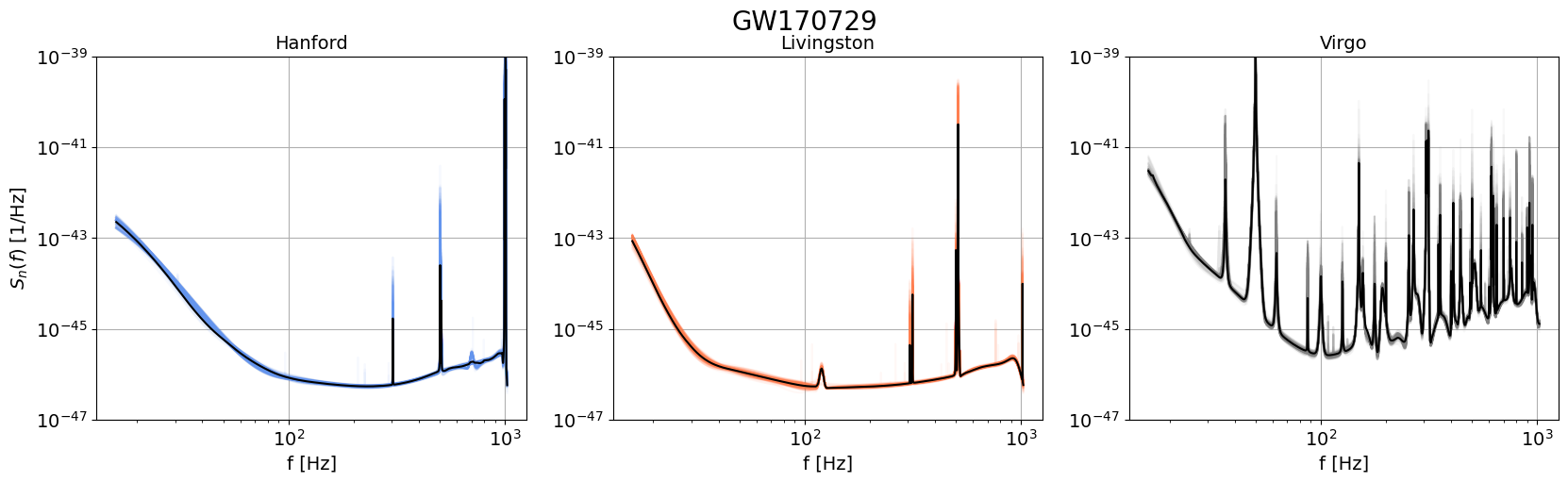}
\includegraphics[width=\textwidth]{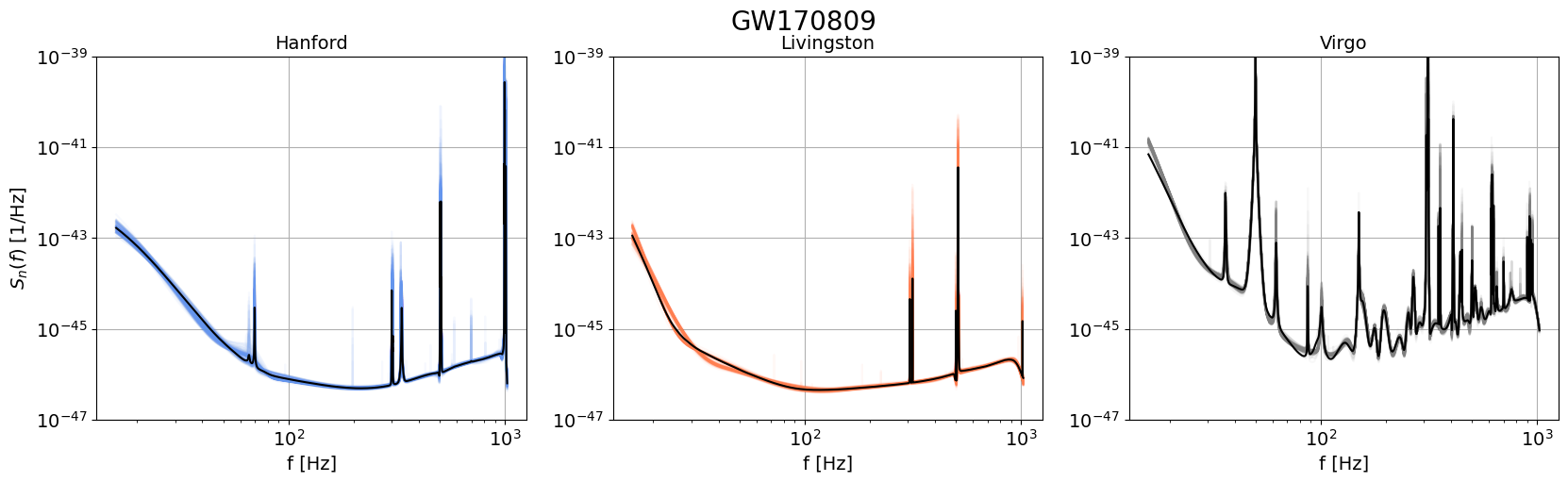}
\caption{\ac{PSD} posterior samples (colored lines) and the median \ac{PSD} (black line) for each of the 11 events in GWTC-1. The \acp{PSD} start at 16~Hz, and the maximum frequency for each event is given in Table.~\ref{tab:settings}.}
\label{fig:psds2}
\end{figure*}

\begin{figure*}
\includegraphics[width=\textwidth]{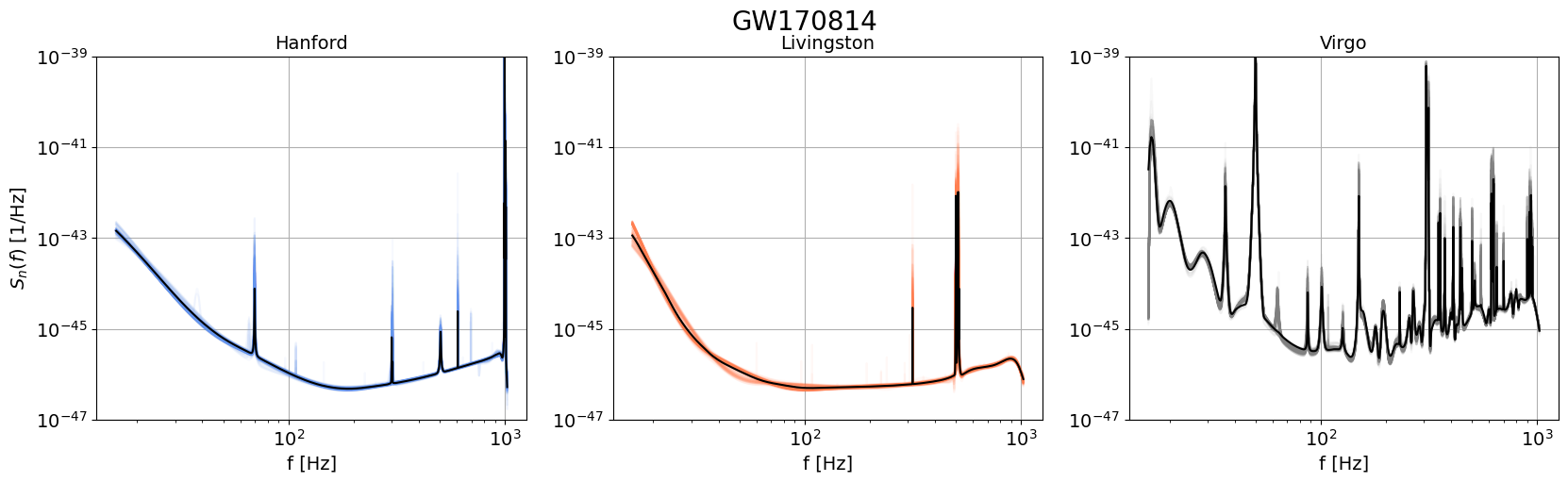}
\includegraphics[width=\textwidth]{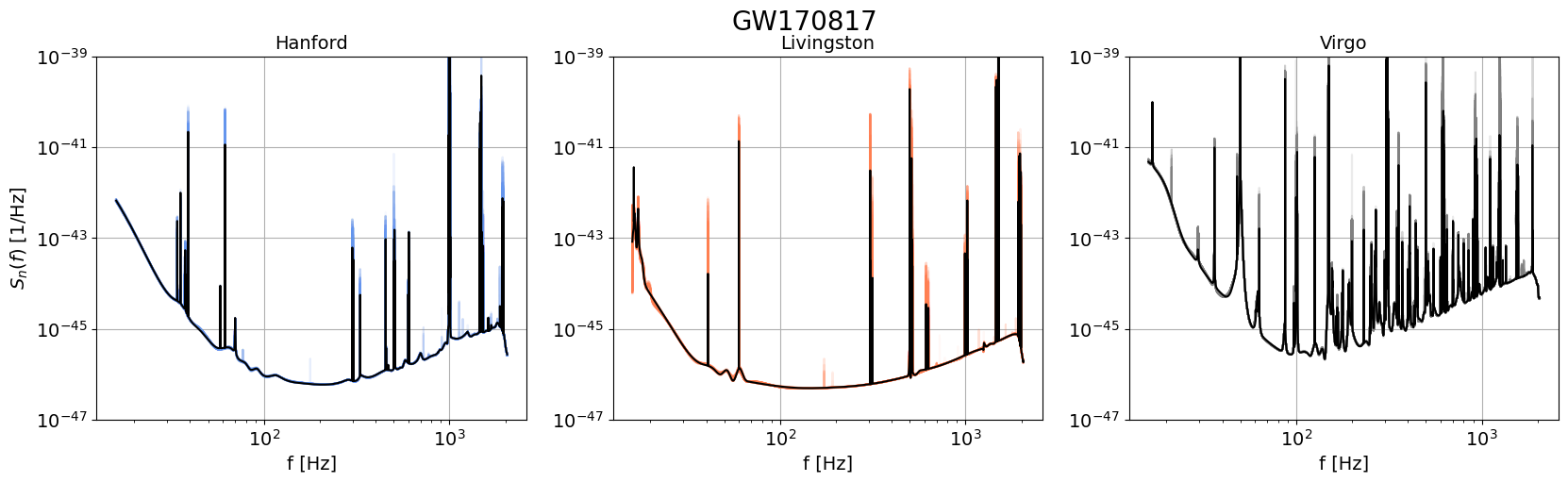}
\includegraphics[width=\textwidth]{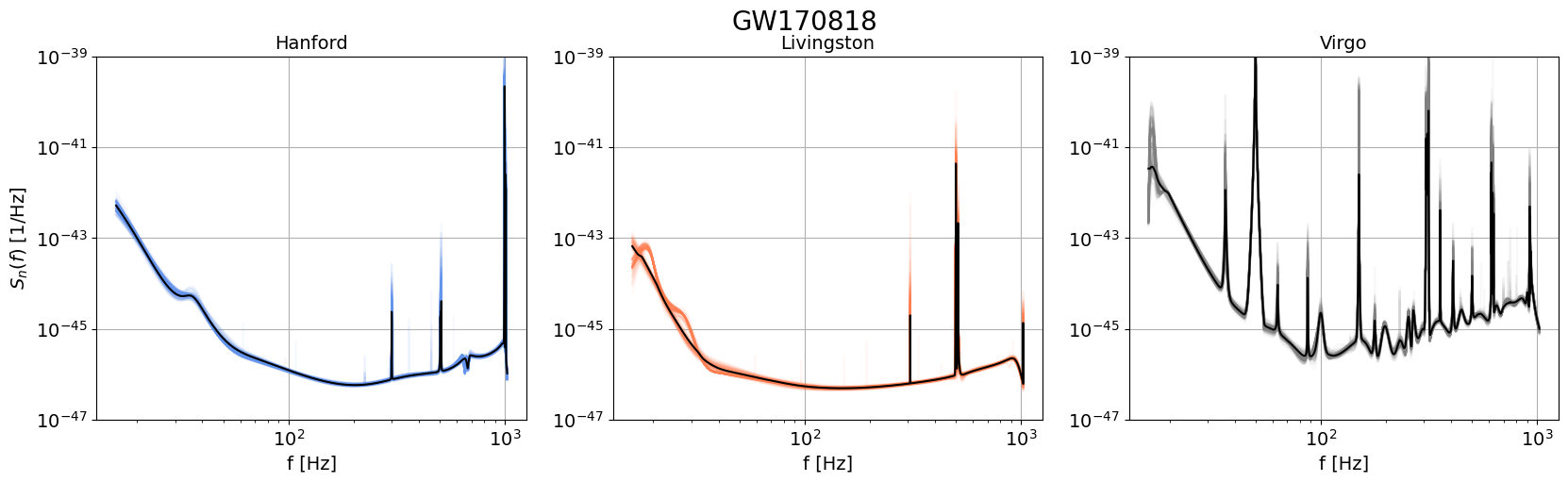}
\includegraphics[width=0.66\textwidth]{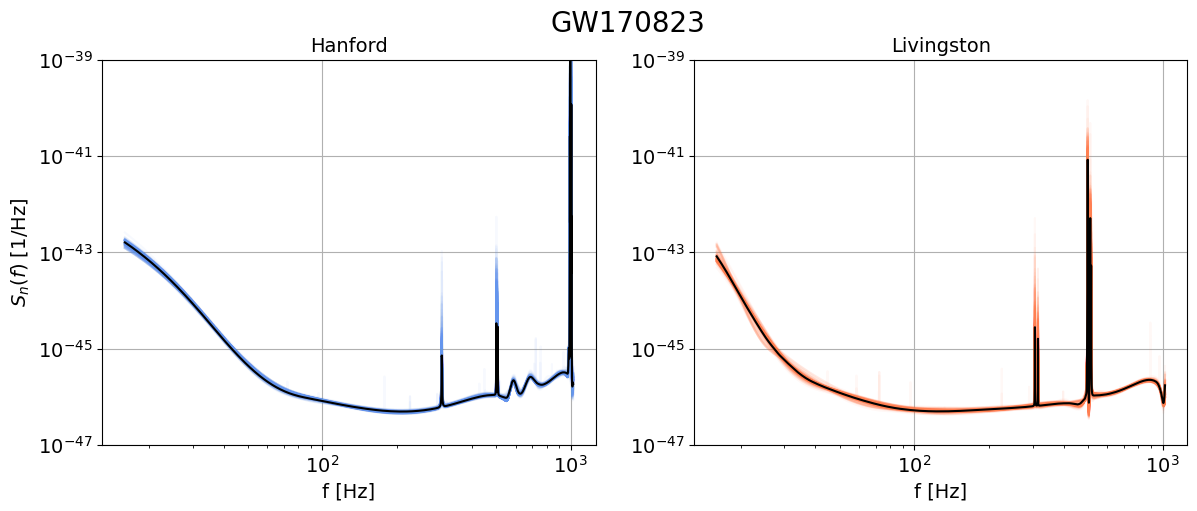}
\caption{\ac{PSD} posterior samples (colored lines) and the median \ac{PSD} (black line) for each of the 11 events in GWTC-1. The \acp{PSD} start at 16~Hz, and the maximum frequency for each event is given in Table.~\ref{tab:settings}.}
\label{fig:psds}
\end{figure*}

In Figs.~\ref{fig:psds1}-\ref{fig:psds}, we present the posteriors for the \acp{PSD} for each detector, LIGO--Hanford and LIGO--Livingston as well as Virgo where such data were available, for each event as well as the resulting median \ac{PSD}.
\label{sec:appendix}

\bibliography{Bibliography}
\end{document}